\documentclass[reprint, aps,floatfix,showpacs,superscriptaddress]{revtex4-1}
\usepackage{graphicx,epsf,epsfig}
\usepackage{subfigure}
\usepackage{color}
\usepackage{float}
\usepackage{dcolumn}
\usepackage{bm}
\usepackage{ulem}
\usepackage{setspace}

\begin{document}

\title{Manifestation of chiral tunneling at a tilted graphene $pn$ junction}

\author{Redwan N. Sajjad}
\affiliation{Department of Electrical and Computer Engineering, University of Virginia, Charlottesville,VA-22904.}
\author{S. Sutar}
\affiliation{College of Nanoscale Science and Engineering, the State University of New York, Albany, NY-12203.}
\author{J. U. Lee}
\affiliation{College of Nanoscale Science and Engineering, the State University of New York, Albany, NY-12203.}
\author{Avik W. Ghosh}
\affiliation{Department of Electrical and Computer Engineering, University of Virginia, Charlottesville,VA-22904.}
\date{\today}


\begin{abstract}
Electrons in graphene follow unconventional trajectories at PN junctions,
driven by their pseudospintronic 
degree of freedom. Significant is the prominent angular dependence of 
transmission, capturing the chiral nature of the electrons and culminating 
in unit transmission at normal incidence (Klein tunneling). We theoretically show 
that such chiral tunneling can be directly observed from the junction 
resistance of a tilted interface probed with separate split gates. The junction 
resistance is shown to increase 
with tilt in agreement with recent experimental evidence. The tilt dependence arises because of the misalignment between modal 
density and the anisotropic transmission lobe oriented perpendicular to the tilt. 
A critical determinant is the presence of edge scattering events that can completely reverse the angle-dependence. The absence of such reversals in the experiments indicates that these edge effects are not overwhelmingly
deleterious, making the premise of transport governed by electron `optics' 
in graphene an exciting possibility.
\end{abstract}
\maketitle
A striking feature of electron flow in graphene, gated uniformly or electrostatically
`doped' into junctions, is the non-trivial dynamics of its pseudospins arising from 
its orthogonal dimer basis sets. The overall photon like dispersion propels 
electrons along trajectories intuitive of Snell's law, conserving quasi-momentum 
components transverse to the interface. However, the corresponding `Fresnel 
equations' are qualitatively different from their optical counterpart, determined 
by conservation of pseudospins. In particular, graphene electrons are chiral in nature 
meaning the pseudospin components are related to the direction of momentum. This results in perfect transmission at normal incidence \cite{katsnelson_06} regardless of voltage gradients across the junction (Klein tunneling).   For other incident angles, the spinor mismatch leads to a unique angle dependent 
transmission across the junction. Thus, while conventional electronics in 
graphene faces possibly steep challenges \cite{schwierz_10}, the dynamics of pseudospintronics 
can usher in novel concepts such as electronic Veselago lens \cite{ref1}, subthermal switches driven by geometry induced metal-insulator transition \cite{sajjad_11} and Andreev reflections \cite{beenakker_06}.

Despite the exciting physics of chiral electron flow, its measurable signatures have 
so far been sparse and indirect. Signatures of Klein tunneling were seen 
\cite{gorbachev_08, stander_09} in the preferential transmission of normally incident 
carriers predicted in \cite{falko_06}. A more direct measurement was the conductance oscillation in an $npn$ structure \cite{young_09}. The reflection amplitude undergoes a phase shift of $\pi$ at normal incidence under the action of a magnetic field, due to the cyclotron bending of the carriers \cite{shytov_08}. However, the main underlying physics of the angle dependent electron transmission has not been explicitly measured. Neither has there been a proper model that can capture both the quantum mismatch of spinors over the entire doping 
regime, as well as diffusive scattering to explore their robustness to impurity and edge scattering events.
\begin{figure}
\includegraphics[width=3.4in]{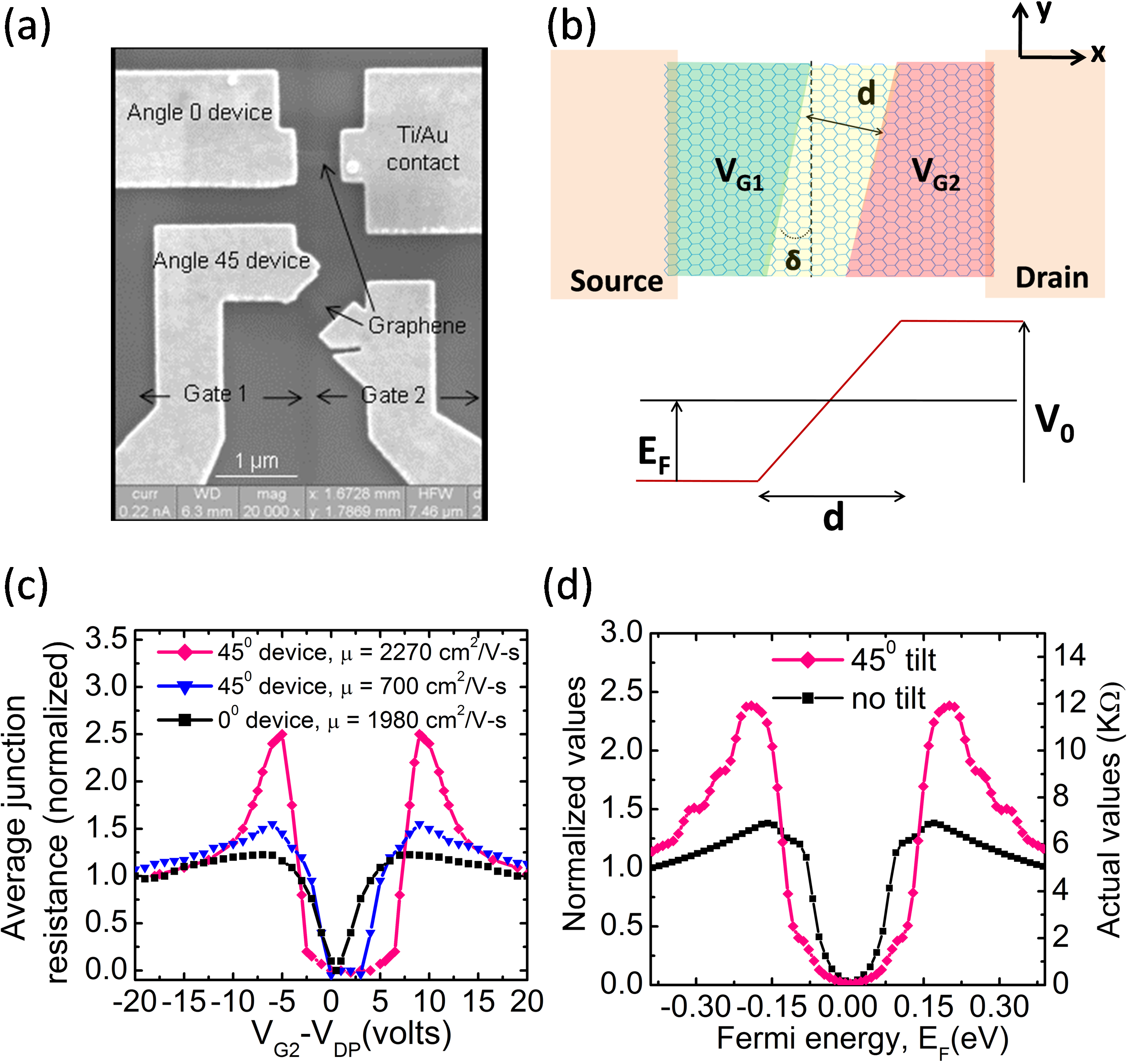}\quad
\caption{(Color online) (a) false color SEM image \cite{sutar_con} and (b) device schematic. Two split gates, separated by a distance $d$ create $pn$ junction with linear potential variation across the junction. The gates are placed at an angle $\delta$ with respect to the transport direction. (c) Experimental \cite{sutar_con}, (d) theoretical junction resistance ($R_J$) vs voltage/energy for $\delta=0,45^0$. In (d), we plot $R_J$ vs $E_F = \hbar v_F \sqrt{\pi \alpha_GC_G|V_{G1}|/q}$ from atomistic NEGF calculation for a 100nm wide graphene sheet.} 
\label{device}
\end{figure}

In this paper we focus on a tilted GPNJ (i) to show that it serves as an explicit signature of chiral tunneling. We show that the junction resistance (similar to the odd resistance shown in other experiments \cite{gorbachev_08,stander_09}) is higher than the non-tilted device (Fig. 1). We argue that this enhancement originates from the chiral nature of graphene electrons which manifests itself through the highly angle dependent transmission characteristics of GPNJ (Fig. 2). The angular transmission lobe, oriented perpendicular to the interface is rotated with the tilt, where fewer transmitting modes exist. Therefore the conductance modulation would not happen for non-chiral, non-relativistic electrons with isotropic transmission. The results follow closely with our recent transport measurements of a tilted GPNJ in a structure that has separately controlled split-gate voltages \cite{sutar_con,sutar_12}. (ii) We present an 
exact analytical solution to the spinor mismatch problem as well as non-equilibrium Green's function (NEGF) based atomistic numerical calculation. An efficient matrix inversion algorithm is employed to reach near experimental dimensions and capture both quantum mechanical and diffusive contributions to the overall resistance. 
We find that charged impurity scattering dilutes, but does not eliminate the modulation in conductance (Figs. 1, 3).  (iii) Notably, we demonstrate that multiple scattering events at edges can reverse the trend in modulation, giving an interfacial resistance that {\it{decreases}} with tilt (Fig. 4). Such decrease has been seen in the past \cite{low_tilted}, but their physical origin has not been identified so far. The absence of such a reversal in experiments surprisingly points to their elimination, possibly through incoherent scattering processes dominant at the strained edges. \\
 {\it{Analytical model.}} The conductance of a GPNJ can be written as, $G = G_0\sum_i^{M} T(\theta_i)$ where $G_0 = 4q^2/h$ is the conductance quantum including spin and valley degeneracies and $T(\theta_i)$ is the incident angle dependent transmission probability with $\theta_i = tan^{-1}({k_y}/{k_x})$. $M$ is the number of modes from the incident side for a given Fermi energy $E_F$ relative to its Dirac point and can be approximated as $M \approx {W}{|E_F|}/{{\pi}\hbar v_F}$ over the linear E-k regime. 
The angle dependent transmission $T$ is obtained by pseudospin conservation across the junction \cite{sajjad_11}
\begin{eqnarray}\label{uni}
T(E_F,\theta_i) &=& \Biggl[\frac{cos\theta_i cos\theta_r}{cos^2\biggl(\displaystyle\frac{\theta_i+\theta_r}{2}\biggr)}\Biggr]e^{\displaystyle -\pi {\hbar v_Fk_{F}^2}dsin^2\theta_i/V_0}\nonumber\\
\end{eqnarray} 
This is a general form of the transmission expression in \cite{low_09, falko_06} and works for the entire voltage range from $PN$ to $NN$ junction. The incident and refracted angles $\theta_i$ and $\theta_r$ are related by  Snell's law \cite{ref1}, $E_F\sin\theta_i = (E_F-V_0)\sin\theta_r$, with $V_0$ (Fig. 1) being the voltage barrier across the junction, and $k_{F}=E_F/\hbar v_F$. The Snell's law arises from transverse quasi-momentum conservation (energy band diagram in Fig. 1(b)). 
For the rest of the paper, we use the average transmission of all modes, $T_{av} = G/M$. 
\begin{figure}
\centering
\includegraphics[width=3.4in]{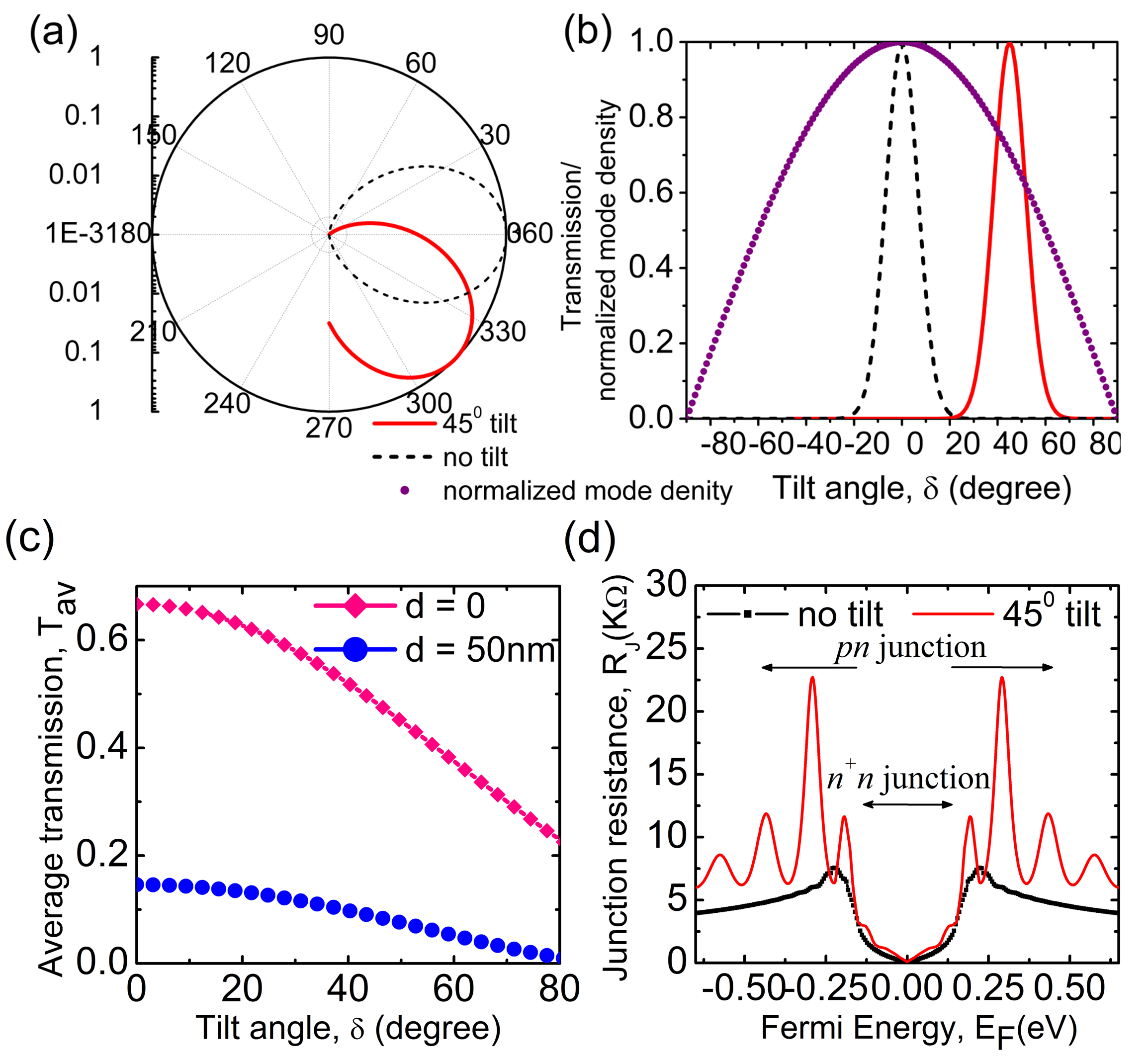}\quad
\caption{(Color online) (a) Angular transmission for various tilt angles. (b) With tilt, the transmission lobe moves into a low mode density ($\sim cos\theta$) area giving (c) a gradual decrease in transmission (Eq. \ref{delta}) for a symmetric $pn$ junction, (d) junction resistance at 0K predicted from Eq. \ref{eq_rj}. We see resonances which become more pronounced as we go into smaller systems.}
\label{tr_tilt}
\end{figure}

For a tilted junction, the incoming mode angles are modified, so that the conductance becomes,
\begin{eqnarray}\label{eq_1}
G(E_F) = G_0\sum_i^{M(E_F)} T(E_F,\theta_i+\delta)
\end{eqnarray}where $\delta$ is the tilt angle as shown in Fig. 1. In Eq. \ref{eq_1} the effective split between the two gates is $d/cos\delta$. As a result of the angle dependence, the 
transmission lobe at a particular energy will  rotate by the tilt angle (Fig. \ref{tr_tilt}a). The transverse wavevector $k_y = ksin\theta$ gives $\Delta \theta = \Delta k_y/Kcos\theta$, so that the mode density decreases as we go to higher angles relative to the {\it{transport axis}}. A tilt at the junction thus shifts the transmission window onto a high angle region where the mode density is less, decreasing the overall transmission  (Fig. \ref{tr_tilt}c). In the limit when the no. of modes is very few, the experiment will give the mode resolved angle dependent transmission properties (Fig. 2(a)). 
For an abrupt, symmetric $pn$ junction, the transmission expression reduces to $cos^2\theta$ from Eq. \ref{uni} and it is easy to see the impact of tilt,
\begin{eqnarray}\label{delta}
G &\approx& G_0\int_{-\pi/2}^{\pi/2-\delta} \frac{T(\theta+\delta)}{\Delta \theta}d\theta = \frac{2}{3}cos^4(\frac{\delta}{2}) M
\end{eqnarray} The factor $2/3$ arises from the wavefunction mismatch across the junction and the tilt
introduces an additional scaling factor, which further reduces conductance. The gradual decrease of $T_{av}$ with $\delta$ in Fig. \ref{tr_tilt}(c) constitutes a direct manifestation of chiral tunneling in graphene. 

To connect with experimental measurements, we next analyze the variation of the junction resistance in presence of an intrinsic background doping ($V_{DP}$) in the graphene sheet (Fig. 2d).  We vary the gate voltages so that $V_{G1}=-V_{G2}$ but a nonzero $V_{DP}$ makes it an asymmetric GPNJ. 
The effective gate voltages on the graphene sheet are $\alpha_G(V_{G1}+V_{DP})$ and $\alpha_G(V_{G2}+V_{DP})$, where $\alpha_G$ is the capacitive gate transfer factor. The junction resistance can be written as \cite{datta_97},
\begin{eqnarray}\label{eq_rj}
R_J = (\frac{4q^2}{h})^{-1}[\frac{1-T_{av}}{MT_{av}}]
\end{eqnarray} 
Fig.~\ref{tr_tilt}(d) plots $R_J$ against $E_F=\hbar v_F \sqrt{(\pi)\alpha_G C_G |V_{G1}|/q}$, the amount of shift in the Dirac point by $V_{G1}$, for a 100nm wide graphene sheet with a split gate separation $d = 200nm$. For the voltage range $|V_{G1}|<V_{DP}$, we are in the $n^+n$ regime for positive $V_{DP}$ and $p^+p$ for negative. Under these near homogeneous conditions, the junction resistance predicted by Eqs.~1-4 is small, because the pseudospin states match and there is no WKB tunneling term in Eq. \ref{uni}. When $|V_{G1}|>V_{DP}$, we are in the $pn$ junction regime and resistance jumps to a high value, primarily due to the WKB factor in Eq. \ref{uni} (similar trend was seen in \cite{gorbachev_08,stander_09}). The rate of change in $R_J$ with  $V_{G1}$ is determined by $d$. For the $45^0$ tilted junction, the junction resistance is higher than the non-tilted resistance. We see oscillation in the resistance (Fig. 2(d)) for the single $np$ junction. This is different from the interference oscillation in Ref. \cite{young_09} for the resonant cavity formed in $npn$ strucure. This can be understood from the conductance in the $pn$ junction regime, simplified  as $G(k_F) \approx G_0\sum_i^M\exp({-\pi k_F{d}sin^2\theta_i/2})$ for large $d$. With increasing gate voltage (higher $k_F$) we have more modes ($M$) in the summation with each mode transmitting with exponentially reduced magnitude. The two opposing effects generate a sequence of peaks and valleys and dominate when we sum over a few modes (either with quantization or tilt) at low temperature. 

{\it{Numerical model.}} An atomistic, numerical calculation of the junction resistance is shown in Fig. 1 (d) at 80K temperature. We use NEGF formalism for a 100nm wide graphene sheet with $d=200nm$, close to experimental dimensions (width $\sim$200-300nm). A single $p_z$ orbital basis for each carbon atom is used to compute the Hamiltonian $H$, while the contact self-energies $\Sigma_{1,2}$ are calculated using a recursive technique. The retarded Green's functions are calculated as 
$G^R = (E_FI-H+V-\Sigma_1-\Sigma_2)^{-1}$ using the algorithm in \cite{anantram_02}, $V$ is the electrostatic potential inside the device.  In units of $ {2q^2}/{h}$, the conductance is calculated as $G =\Gamma_1G^R\Gamma_2G^A$, where the contact broadening functions $\Gamma$ are the anti-Hermitian components of $\Sigma$. For ballistic transport,
$G$ equals $M$, the number of modes for a uniformly gated sheet, and $MT_{av}$ for a $pn$ junction. Combining these two, we calculate $T_{av}$ and $R_J$ from Eq. \ref{eq_rj}, which shows a jump with tilt in the $pn$ junction regime, very similar to the experiment.

{\it{Charged impurity scattering dilutes tilt dependence.}}
The experimental device is on a SiO$_2$ substrate and the transport is diffusive with a mobility varying from 700-3000 cm$^2$/V-s. It is natural to inquire how the theoretical model, which so far had no scattering, corresponds to experiments. To explore this feature, we included the impact of charged impurity scattering in our model. We use a sequence of screened Gaussian potential profiles for 
\begin{figure}
\centering
\includegraphics[width=3.4in]{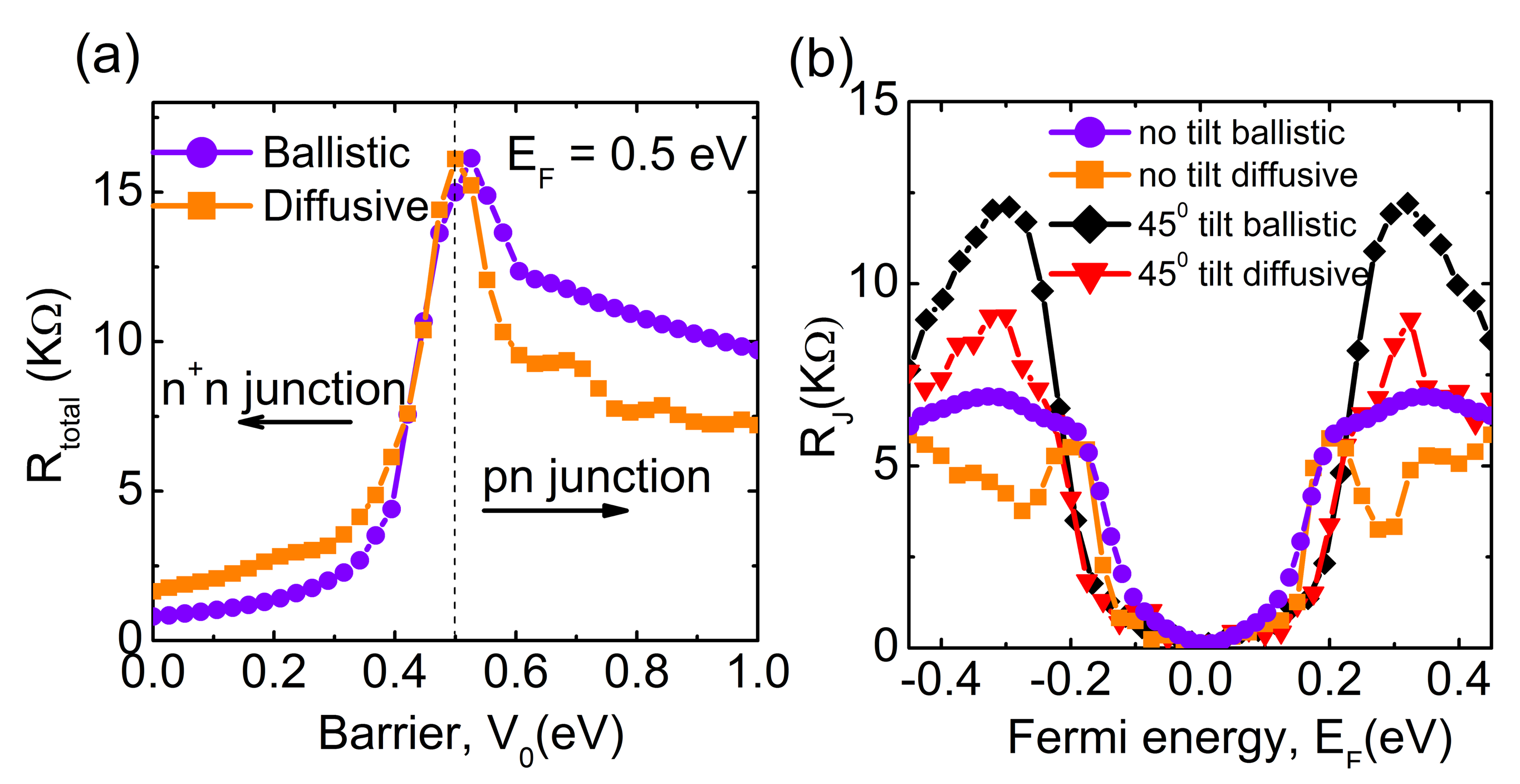}\quad
\caption{(Color online) Impact of charged impurity scattering, (a) conductance asymmetry is diluted due to impurity potentials, the ballistic resistance is normalized for comparison. (b) Reduced asymmetry results in lower junction resistance for both tilted, non-tilted devices, thus retaining their difference.} 
\label{device}
\end{figure}the impurity scattering centers \cite{klos_10, peres_10, lewenkopf_08}, 
$U(r) = \sum_{n=1}^{N_{imp}} U_n\exp{(-{|r-r_n|^2}/{2\zeta^2})}$, that specifies the strength of the impurity potential at atomic site r, with ${r_n}$ being the positions of the impurity atoms and $\zeta$ the screening length ($\approx$ 8 times the C-C bond for long range scatterers). The amplitudes $U_n$ lie in the range $[-\delta, \delta]$ ($\approx$ 0.5 times the C-C coupling parameter) and $N_{imp}$ is the impurity concentration ($\sim$ 5 x 10$^{11}$ cm$^{-2}$). Note that the purely diffusive model discussed in \cite{fogler_08,stander_09,gorbachev_08} ignores the quantum mechanical spinor mismatch and WKB scaling and therefore underestimates the junction resistance for cleaner samples. Our NEGF based numerical model on the other hand captures both the quantum mechanical and impurity limited resistance contributions simulataneously. The junction resistance is now calculated by eliminating the contact and device resistance \cite{sutar_12}
\begin{eqnarray}
R_J = [R(V_{G1},V_{G2})+R(V_{G2},V_{G1}) \nonumber\\
-R(V_{G1},V_{G1})-R(V_{G2},V_{G2}]/{2},
\end{eqnarray}where the first two terms contain the junction resistance, while the last two don't.

 Fig. 3(a) shows the impact of the impurity scatterings on the total resistance and 3(b) on the junction resistance. We take the average resistance over many impurity configurations. This puts a constraint on the computation size, so  we show calculations this time for a smaller device (50nm wide). We find that both tilted and non-tilted junction resistances are suppressed, thereby retaining the difference between the two. This reduction in junction resistance with scattering is quite consistent with the experiment (Fig. 1(c), red line is for a $45^0$ device with mobility 2270cm$^2$/V-s, while the blue triangle is for $45^0$ with lower mobility, 700cm$^2$/V-s). 

The reduction in the junction resistance from ballistic to diffusive transport can be understood from the trend in total resistance, shown in Fig. \ref{device}(a). Now we keep $V_{G1}$ fixed and vary $V_{G2}$ so that we go from $n^+n$ to $np$ junction. We see a clear asymmetry in the R-V$_G$ \cite{low_09, stander_09} (purple line normalized to the orange line in Fig. 3(a) for comparison). The asymmetry confirms the presence of $pn$ junction, which reduces the conductance due to spinor mismatch. The presence of impurity scattering reduces this asymmetry while increasing the overall resistance (red line). The impurity potentials create a random potential variation throughout the graphene sheet on top of the applied gate voltages, thus blurring the presence of a $pn$ junction. Therefore the resistance due to spinor mismatch becomes less noticeable (Fig. 3(b)). Indeed, the experimental data of the total resistance indicates an increase in asymmetry in the tilted junction \cite{sutar_12}, signifying an increase in the junction resistance.

\begin{figure}
\centering
\includegraphics[width=3.2in]{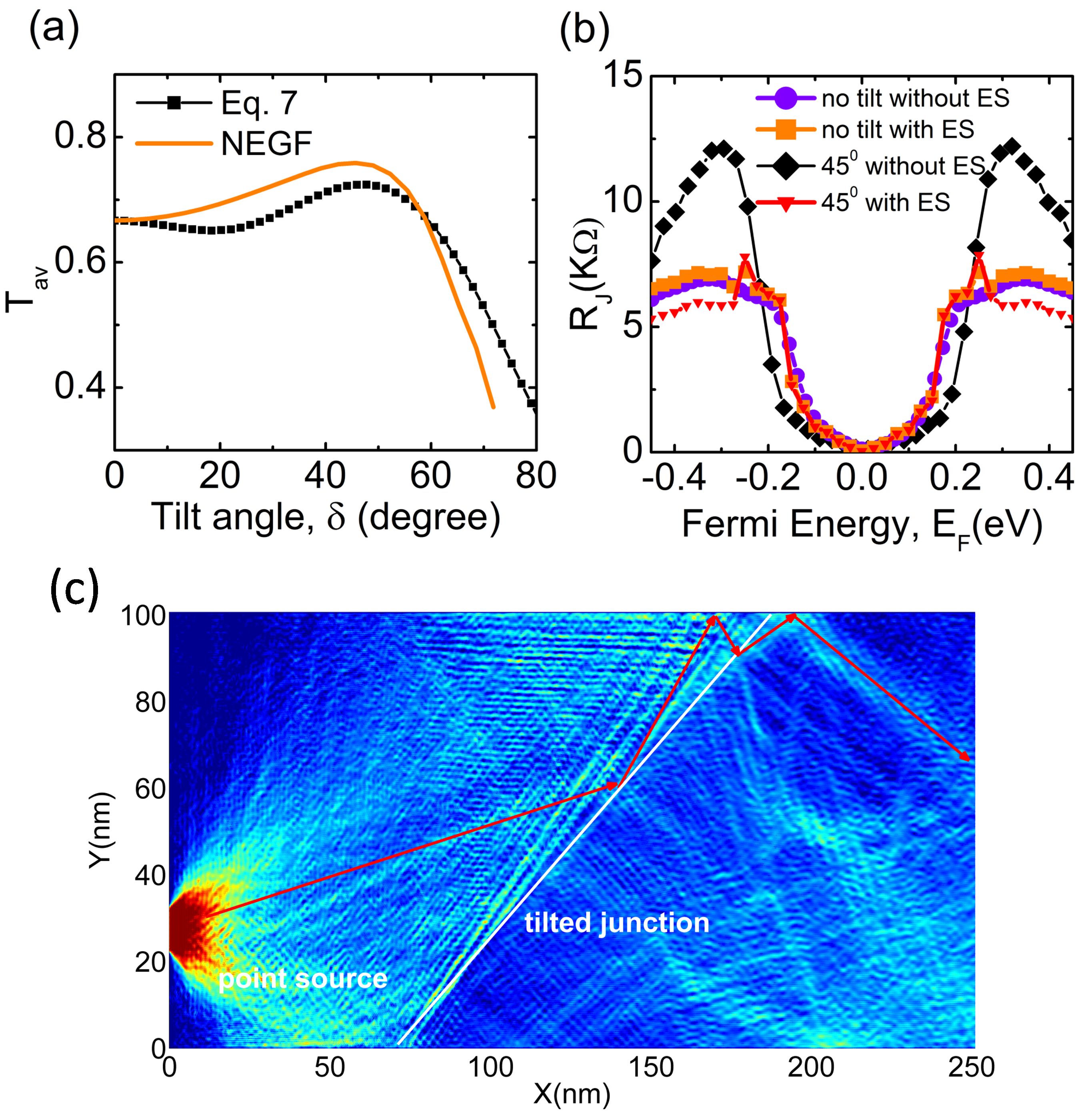}\quad
\caption{(Color online) (a) Increase in conductance in a tilted GPNJ due to Edge Scattering (ES) in contrast with Fig. 2(c). (b) corresponding decrease in junction resistance due to tilt (c) Mechanism of edge enhanced conductance for a tilted junction from atomistic NEGF calculation: some reflected electrons come back at the junction after edge reflection. 
}
\label{color}
\end{figure}

{\it{Edge scattering can reverse tilt dependence.}} A striking feature on the experimental results is their agreement with Eq. \ref{delta}. This match is remarkable, considering that the equation was derived assuming no edge reflections and the fact that past numerical study \cite{low_tilted} showed in fact an {\it{increase}} in conductance with tilt. We argue that the above reversal of junction conductance with tilt is entirely due to edge scattering events. Indeed, from an atomistic NEGF calculation with shorter widths than lengths, we find that the transmission now shows a pronounced local maximum (Fig.~\ref{color}(a) orange line) in agreement with \cite{low_tilted}, increasing thereby the junction resistance.
We summarize this in Fig.~\ref{color}(b),  where an increasing tilt makes the resistance increase for the short channel 125nm x 50nm device (a transition from purple circle to black diamond), but decrease for the longer 200nm x 50nm device (orange square to red triangle). Bearing in mind that the gate split is 100nm, the short channel device significantly reduces edge scatterings.

To better understand the origin of such a resistance reversal, we inject electrons with a small contact at the 
left edge (bright red spot in Fig.~\ref{color}c) and plot the spatial current density under a small drain bias. The numerically computed electron trajectories show how a tilt can enhance forward scattering events at the edge and thus an increase in conductance. The enhancement arises from simple `geometrical optics' dictated by Snell's law. We can identify the incident wide angle modes ($\theta>\pi/4-\delta$), for which the reflected `ray' hits the upper edge with a positive $x$ directed velocity. Such a mode will reflect back towards the junction again. The contribution from the positively directed edge scattering event is given by
\begin{eqnarray}
G_{edge}=G_0\int_{\pi/2-3\delta}^{\pi/2-\delta} \frac{T(\pi-3\delta-\theta)}{\Delta \theta}d\theta = 2\sin^4\delta \cos\delta M\nonumber\\
\end{eqnarray}Note that only the incident angles below the critical angle are considered while setting the limits of the integration.

With the added contribution from edge scattering, the net mode-averaged transmission $T = G/M$ is given by
\begin{equation}\label{delta_edge}
T_{total} \approx T + (1-T)T_{edge}\eta 
\end{equation}
where the $T$s are extracted from the corresponding $G/M$ ratios and $\eta$ is a parameter that describes the efficiency of edge scattering. In the absence of edge scattering ($\eta=0$), $T_{total} = T$ and decreases with tilt (Fig. 2c). However in the presence of strong edge scattering ($\eta = 1$), the added forward edge scattering term in Eq. \ref{delta_edge} closely reproduces the NEGF result with the local transmission maximum (Fig.~\ref{color}a, black dotted line). Comparing these results with experiments indicates that such edge scattering events are clearly minor. We conjecture that the coherent forward scattering processes captured by NEGF can be diluted down in the experiments by the presence of incoherent scattering processes arising at the strained and rough edges of the graphene samples that tend to dephase or perhaps even trap the electrons. 

The experimental observation of chiral tunneling, particularly in the face of impurity and edge scattering, opens up the possibility of graphene's `geometric optics' based applications, e.g. lens, switches and wave-guides. Analogous results are expected  in bilayer graphene, but not in achiral materials like 2-D hexagonal boron nitride.  

The authors acknowledge funding support from INDEX-NRI and useful discussions with T. Low.

\bibliographystyle{apsrev4-1}

\end{document}